\documentclass[10pt,conference]{IEEEtran}
\usepackage[numbers]{natbib}

\usepackage{cite}
\usepackage{amsmath,amssymb,amsfonts}
\usepackage{algorithmic}
\usepackage{graphicx}
\usepackage{textcomp}
\usepackage{subfigure}
\usepackage{xcolor}
\usepackage{tikz}
\usetikzlibrary{quantikz}
\usepackage{adjustbox}
\usepackage{subfigure}

\usepackage{dcolumn}
\usepackage{bm}
\usepackage{xcolor}
\usepackage[colorlinks=true,linkcolor=purple,citecolor=purple]{hyperref}%

\begin{document}

\title{Best practices for quantum error mitigation with digital zero-noise extrapolation}

\author{
\IEEEauthorblockN{
Ritajit Majumdar\IEEEauthorrefmark{7}\IEEEauthorrefmark{5}, 
Pedro Rivero\IEEEauthorrefmark{1}\IEEEauthorrefmark{5},
Friederike Metz\IEEEauthorrefmark{1}\IEEEauthorrefmark{2}\IEEEauthorrefmark{3},
Areeq Hasan\IEEEauthorrefmark{1}\IEEEauthorrefmark{4}, and
Derek S. Wang\IEEEauthorrefmark{1}\IEEEauthorrefmark{6}
}

\IEEEauthorblockA{\IEEEauthorrefmark{7} \textit{IBM Quantum}, IBM India Research Lab}
\IEEEauthorblockA{\IEEEauthorrefmark{1} \textit{IBM Quantum}, IBM T. J. Watson Research Center}
\IEEEauthorblockA{\IEEEauthorrefmark{2} \textit{Institute of Physics}, École Polytechnique Fédérale de Lausanne}
\IEEEauthorblockA{\IEEEauthorrefmark{3} \textit{Center for Quantum Science and Engineering}, École Polytechnique Fédérale de Lausanne}
\IEEEauthorblockA{\IEEEauthorrefmark{4} \textit{Department of Electrical and Computer Engineering}, Princeton University}
\IEEEauthorblockA{\IEEEauthorrefmark{5} Equal contribution}
\IEEEauthorblockA{\IEEEauthorrefmark{6} Email: derek.wang@ibm.com}
}

\maketitle
\thispagestyle{plain}
\pagestyle{plain}

\begin{IEEEkeywords}
zero-noise extrapolation, gate folding, identity insertion, quantum error mitigation, quantum software, quantum applications
\end{IEEEkeywords}

\begin{abstract}
Digital zero-noise extrapolation (dZNE) has emerged as a common approach for quantum error mitigation (QEM) due to its conceptual simplicity, accessibility, and resource efficiency. In practice, however, properly applying dZNE to extend the computational reach of noisy quantum processors is rife with subtleties. Here, based on literature review and original experiments on noisy simulators and real quantum hardware, we define best practices for QEM with dZNE for each step of the workflow, including noise amplification, execution on the quantum device, extrapolation to the zero-noise limit, and composition with other QEM methods. We anticipate that this effort to establish best practices for dZNE will be extended to other QEM methods, leading to more reproducible and rigorous calculations on noisy quantum hardware.
\end{abstract}

\section{Introduction}

Today, quantum computers are too noisy to run fault-tolerant error correcting codes, so quantum error mitigation (QEM) methods are necessary for extending the reach of small, noisy quantum devices~\citep{Bravyi2022, Kandala2017}. Such methods typically involve running additional or modified mitigation circuits in addition to or instead of the target circuits. Therefore, the use of QEM techniques generally results in longer execution times or requires access to additional qubits for increased accuracy. An example of the former is probabilistic error cancellation \citep{Temme2016, vandenBerg2023} where the noise of the target circuit is learned layer by layer and then cancelled in a probabilistic manner with an exponential overhead to control the subsequent spread in the variance of expectation values. An example of the latter is virtual distillation that requires high connectivity to `distill’ multiple copies of the noisy quantum state into a pure one~\citep{Huo2021, Huggins2021, GoogleQEMReview2022}.

\begin{figure}[ht]
\centering
\includegraphics[width=0.95\linewidth]{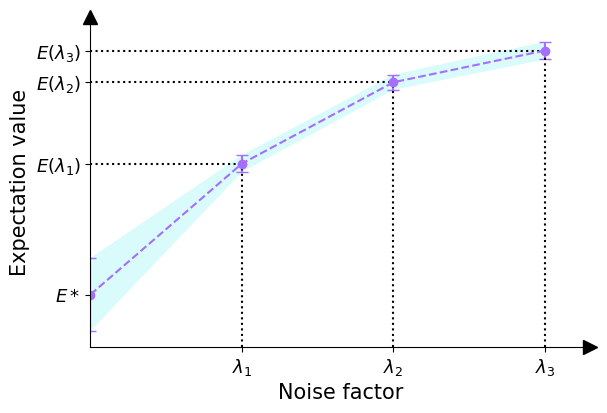}
\caption{Zero-noise extrapolation is a quantum error mitigation technique, where the zero-noise result $E^*$ of a quantum computation is estimated by extrapolating from the result at multiple different noise levels $[\lambda_1, \lambda_2, \lambda_3, ...]$.
} 
\label{fig:zne}
\end{figure}

One of the most common methods for QEM that involves running additional mitigation circuits is zero-noise extrapolation (ZNE) \citep{Temme2016, PhysRevX.7.021050}, as shown in Fig.~\ref{fig:zne}.
In digital ZNE (dZNE)~\citep{GiurgicaTiron2020ZNEReviewAndAdaptive}, the ``zero-noise" expectation value of the target circuit is extrapolated from the expectation values of mitigation circuits where the noise has been amplified by inserting additional digital quantum gates. In this work, we focus on the insertion of identities due to the simplicity of implementation, although quasi-probability-based approaches are possible as well~\citep{Mari2021PEA, Kim2023Utility}. Importantly, this method can be resource-efficient in the low-noise limit by invoking moderate overhead. Error bounds, however, are less certain given the rather liberal assumptions about the underlying device noise. In addition, it is especially accessible to application developers compared to analog zero-noise extrapolation (aZNE) where noise is tuned at the hardware-level by, \textit{e.g.}, stretching and scaling the pulses themselves. aZNE therefore requires deep knowledge of and access to the underlying device physics~\citep{Kim2023}. Given these advantages of dZNE, several hardware-agnostic implementations exist~\citep{Rivero2022ZNE, LaRose2022} that have been leveraged in previous scientific works~\citep{UnitaryBenchmarkQEM2022, GoogleQEMReview2022, Kirmani2023Braiding}, suggesting dZNE is the industry standard for QEM.

\begin{figure*}[ht]
\centering
\includegraphics[width=1.0\linewidth]{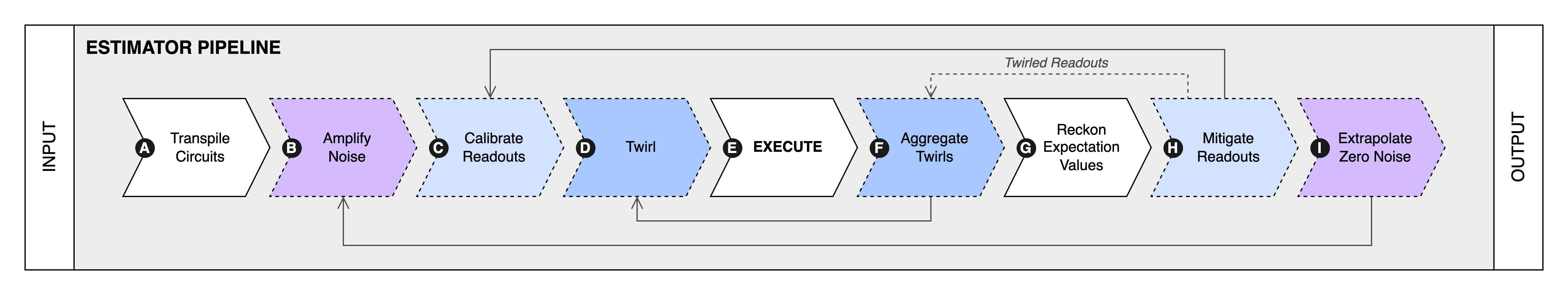}
\caption{Schematic of an error-mitigated \texttt{Estimator} pipeline \citep{Qiskit, Rivero2023StagedPrimitives} for noisy quantum computations that takes quantum circuits as inputs and returns expectation values. Depicted QEM methods include digital zero-noise extrapolation (stages $\{B,I\}$), readout error mitigation (stages $\{C,H\}$), and Pauli twirling (stages $\{D,F\}$). 
Stages $\{A,E,G\}$ comprise the unmitigated base pipeline to which additional mitigation techniques can be injected. 
Arrows signify functional requirements, while dotted arrows signify partial or optional dependencies.
} 
\label{fig:schematic}
\end{figure*}

While dZNE can be conceptually straightforward, there are myriad subtleties in its practical application. Therefore, in this work, we suggest best practices for dZNE. We first outline a general and flexible workflow for dZNE in Fig.~\ref{fig:schematic}, and then we evaluate how to best configure each step within this framework. These steps include noise amplification, execution on quantum hardware, and extrapolation to the zero-noise expectation value. We also discuss ways to augment this pipeline to mitigate other sources of noise, such as measurement and coherent gate errors. 

We benchmark with circuits for spin dynamics and run calculations on both noisy quantum simulators and real quantum hardware as concrete examples. From these examples, we find, for instance, that multiple noise-amplified circuits should be sampled for partial noise factors; the noise factors and extrapolator can be chosen based on a calibration; and that ZNE fails for flat noise profiles when noise saturates given practical limits on shots. Notably, we emphasize that these conclusions are highly context-dependent---our conclusions are a good starting point and expected to be widely applicable, but they are subject to change based on differences in the device noise model, the structure of the target circuit, the software implementation, and advancements in the theory of dZNE. Nonetheless, we anticipate that this effort to establish best practices for dZNE will enable more reproducible and higher-quality computations on noisy quantum devices, as well as inspire similar efforts for other QEM approaches.

\section{Noise amplification} \label{noiseamplification}

In the most popular implementation of dZNE, we assume that the noise can be amplified \textit{digitally} by ``folding" each gate to reach a desired noise level. For instance, in the ideal case, the noise of a target circuit $\hat{U}$ would be amplified threefold in $\hat{U}\hat{U}^\dagger\hat{U}$, where $\hat{U}^\dagger \hat{U}$ resolve to identity. Amplifying the noise digitally in this fashion, however, requires several assumptions to hold true, in particular that $\hat{U}$ and $\hat{U}^\dagger$ contain the same levels of incoherent noise. (We note that amplifying noisy hardware-native gates is especially convenient on IBM's superconducting devices because the noisiest gate, \texttt{CNOT}, is its own inverse, although more precise pulse-level inverses that require some calibration can enable even more accurate noise amplification~\citep{henao2023adaptive}.) Therefore, in this section, we discuss how best to implement dZNE to adhere to these assumptions as closely as possible.

\textit{Transpile first.} First, noise should be amplified in circuits that have already been transpiled to a particular device, where transpilation (or ``quantum compilation") refers to the process of preparing a circuit to run on actual quantum hardware \citep{Maronese2021QuantumCompiling}. These steps include mapping logical qubits to device qubits based on connectivity requirements and decomposing the gates into hardware-native digital gates. If amplification were to occur before transpilation, then it is possible that folding would under- or over-represent the noise, leading to a potential bias in the extrapolation process. It could also be the case that $\hat{U}$ and $\hat{U}^\dagger$, perhaps due to stochastic algorithms commonly used for transpilation \citep{li2019tackling}, result in circuits with drastically different structures and, therefore, noise profiles.

\begin{figure}[ht]
\centering
\includegraphics[width=1.0\linewidth]{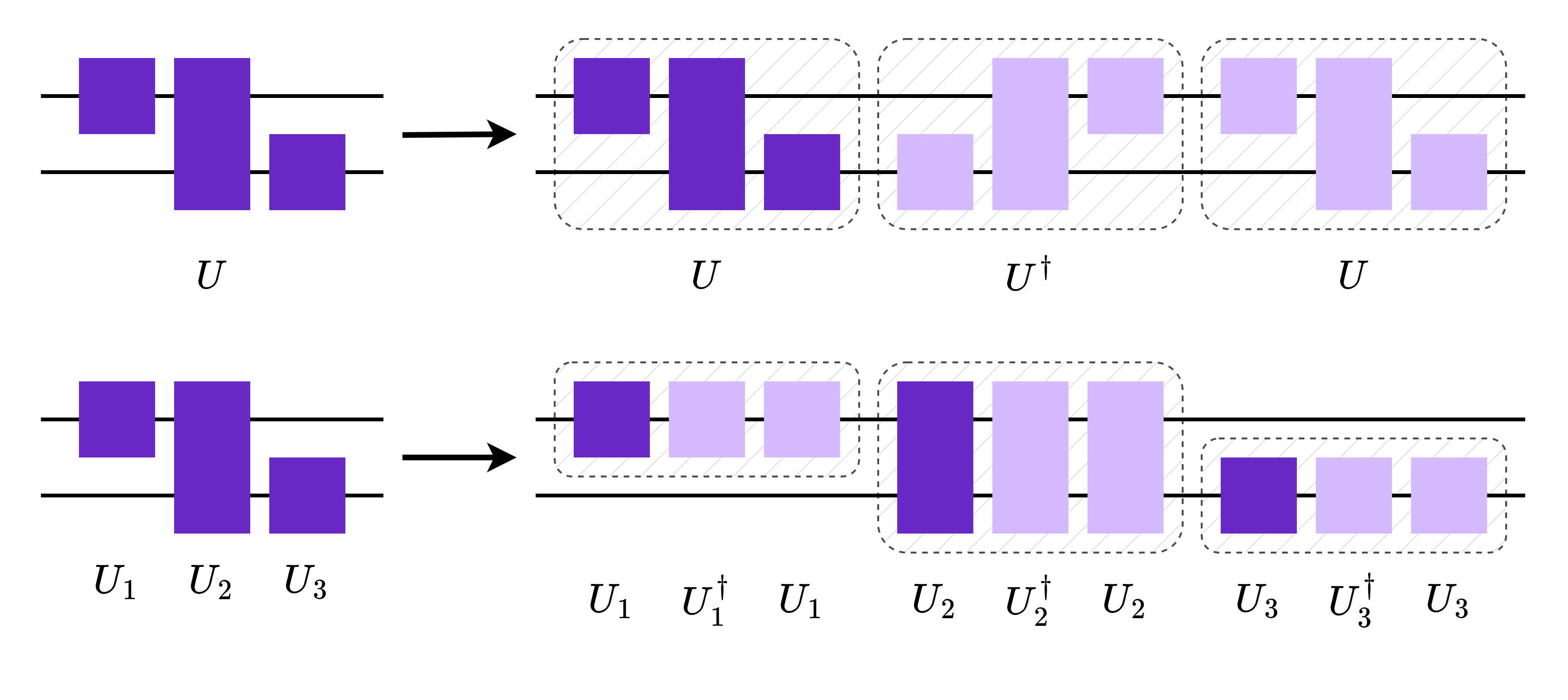}
\caption{Global (upper) vs. local (lower) folding for digital noise amplification.
} 
\label{fig:localglobalfolding}
\end{figure}

\textit{Local and global folding.} Next is the question of folding globally or locally. For a target circuit $\hat{U}=\hat{U}_1\hat{U}_2\hat{U}_3$, we refer to global folding (to a noise factor of 3 in this case) as  $\hat{U}\hat{U}^\dagger\hat{U}=(\hat{U}_1 \hat{U}_2 \hat{U}_3) (\hat{U}_3^\dagger \hat{U}_2^\dagger \hat{U}_1^\dagger) (\hat{U}_1 \hat{U}_2 \hat{U}_3)$ and local folding as $(\hat{U}_1 \hat{U}_1^\dagger \hat{U}_1) (\hat{U}_2 \hat{U}_2^\dagger \hat{U}_2) (\hat{U}_3 \hat{U}_3^\dagger \hat{U}_3)$; see Fig.~\ref{fig:localglobalfolding}. For simple noise models, such as global depolarizing noise, these two approaches should result in identical noise amplification. However, in the presence of more complex noise, this is not necessarily the case. For instance, in Ref.~\citep{Schultz2022LocalGlobalFoldingTimeCorrelatedNoise}, the authors find that global folding amplifies the noise more reliably than local folding in the presence of time-correlated noise. While global folding confers resistance to time-correlated noise, it is conceptually more straightforward to locally amplify the noise to factors that are not odd, positive integers in $\{1, 3, 5, ..., 2n+1\}$. Therefore, within this manuscript, we locally fold circuits.

\begin{figure}[ht]
\centering
\includegraphics[width=0.9\linewidth]{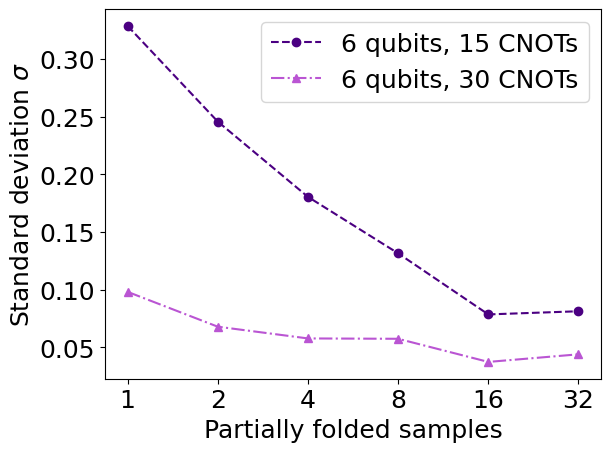}
\caption{Partial folding should be executed with care. Because not all gates are folded, we suggest to sample multiple partially folded circuits at each noise factor. In this example, for a circuit over 6 qubits with either 15 or 30 CNOT gates whose errors are different for each qubit and range between 1\% to 10\%, we find that increasing the number of circuits randomly sampled at a noise factor of 1 and 1.1 decreases the standard deviation $\sigma$ of the zero-noise expectation value $\langle Z_0 Z_1 Z_2 Z_3 Z_4 Z_5 \rangle$ for fewer total CNOT.
} 
\label{fig:partialfolding}
\end{figure}

\textit{Partially folding.} Locally amplifying noise to factors that are not $\{1, 3, 5, ..., 2n+1\}$ is also known as partially folding circuits because gates in the original circuit are not all folded an equal number of times. Partially folding is useful for amplifying noise with a finer resolution or to smaller factors when circuit noise is large, as we demonstrate in further detail in Section~\ref{sec:extrapolation}. In this case, the best practices for doing so are less obvious. There have been several previous proposals---for instance, the circuits can be folded globally right-to-left or left-to-right~\citep{GiurgicaTiron2020ZNEReviewAndAdaptive} or randomly~\citep{GiurgicaTiron2020ZNEReviewAndAdaptive, He2020} with weights based on the relative error of each gate~\citep{Pascuzzi2022WeightedFolding, LaRose2022}. 

Although these approaches can account for instances in which noise is amplified differently when gates are folded earlier vs. later in the circuit and in which each gate on each qubit has different errors, respectively, they are less broadly generalizable---for instance, gate calibrations can become stale. Therefore, we suggest to randomly sample \textit{multiple} partially folded circuits at each noise factor~\citep{He2020}. Here, we do so with the Qiskit Prototype for ZNE~\citep{Rivero2022ZNE}, although additional implementations are available, \textit{e.g.} Ref~\citep{LaRose2022}. Doing so incurs a minimal overhead, as while the number of circuits increases, the number of shots per circuit can be decreased proportionally to conserve the overall shot overhead. We demonstrate concretely in Fig.~\ref{fig:partialfolding} that this approach can decrease the variance of expectation value estimates. Here, we run a brickwork circuit of \texttt{CZ} on noisy simulators with depolarizing gate errors between pairs of qubits ranging from 1\% to 10\% with noise factors of 1 and 1.1, and we measure $\langle Z \rangle=\langle Z^{\otimes n} \rangle$; see Appendix \ref{app:experiment} for further details on the circuits. Although the mean expectation value is the same regardless of the number of partially folded samples, we find that sampling multiple partially folded circuits is necessary when proportionally few of the total gates are folded. As the total number of gates increases, we see a comparatively smaller benefit to sampling noise-amplified circuits at each partial noise factor.

\section{Execution} \label{sec:execution}

In a typical quantum computing workflow, one submits a batch of multiple circuits to the device to compute the expectation value of each. Then, to find the zero-noise extrapolated value, one also submits batches of folded circuits to compute the expectation values at various noise factors. 

\textit{Interleaved execution.} One subtle but practically relevant question is the order in which these target and amplified circuits should be run on the device. Given that quantum device noise is subject to change over time, we suggest to \textit{interleave} the folded circuits with the target circuits, as also suggested in Ref.~\citep{henao2023adaptive}, so that observables are computed sequentially.

\begin{figure}[ht]
\centering
\includegraphics[width=1.0\linewidth]{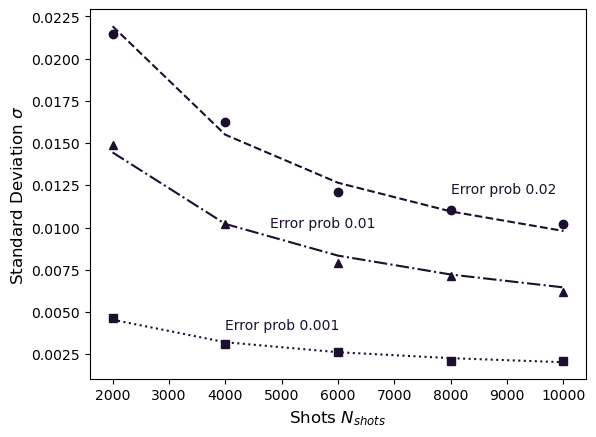}
\caption{The standard deviation of $\langle Z \rangle$ with number of shots scales as $\frac{a}{\sqrt{N}_\mathrm{shots}}$ (fitted curve) for different depolarization error probabilities. Although the order of the scaling remains invariant with the error probability, the scaling factor $a$ increases with the error probability.}
\label{fig:shots}
\end{figure}

\textit{Shots.} Another important consideration is the shot overhead. It is well known that shot noise in the expectation values of Pauli observables without zero-noise extrapolation can be modeled with a binomial distribution and scales as $N_\mathrm{shots}^{-1/2}$. The standard deviation of the zero-noise extrapolated value with a linear extrapolator due to shot noise follows the same relationship, as we show in Fig.~\ref{fig:shots}. Here, $N_\mathrm{shots}$ refers to the number of shots at each noise factor, and we evaluate the standard deviation $\sigma$ for $\langle Z \rangle=\langle Z^{\otimes n} \rangle$ on a 6-qubit circuit for spin dynamics with a 2-qubit depth of 10 and noise factors of 1, 3, and 5. We also show the scaling of the shot noise with the number of shots for different depolarization error probabilities. We observe that in all the cases, the shot noise follows the $N_\mathrm{shots}^{-1/2}$ scaling. 

\section{Extrapolation} \label{sec:extrapolation}

\textit{A priori}, it is challenging to select the best settings for extrapolation, in particular the noise factors and form of the extrapolation function. For instance, consider cases where the exact form of the extrapolation function is even known: exponential extrapolation in the presence of certain incoherent noise \citep{Endo2018, GiurgicaTiron2020ZNEReviewAndAdaptive, Cai2020MultiexponentialEE, henao2023adaptive}. Intuitively, it is desirable to choose noise factors that span the largest possible range without creeping into the noise-saturated regime to minimize the variance in the extrapolated value, but doing so requires some knowledge of the decay rate of the expectation value with the noise factor. To address these issues, there do exist adaptive approaches~\citep{GiurgicaTiron2020ZNEReviewAndAdaptive, henao2023adaptive}. In the former, the shots are dynamically allocated between circuits with a range of noise factors, but they are accompanied by certain assumptions, such as the form of the extrapolation function, and require interactive access to a quantum device. 

Therefore, dZNE is typically executed in a ``batched" manner, where the target and noise-amplified circuits are executed sequentially, so the noise factors and extrapolation function should be carefully chosen beforehand. One approach is to simply guess, and if the uncertainty of the fit is too large or if the resulting expectation does not make sense based on problem-specific knowledge, then the calculation can be re-submitted. This approach, however, does not scale towards larger systems of the imminent future where the ideal values cannot be computed classically for verification.

\textit{Calibration.} A more informed approach is to \textit{calibrate} dZNE by running circuits similar to the target circuits for which the ideal expectation values are known, determining the best settings for zero-noise extrapolation, and using these for dZNE on the target circuit. Ref.~\citep{Lowe2021CDRandZNE} introduces such a method, where target circuits with gates that make the target non-trivial to simulate on classical computers are transformed into classically simulable Clifford circuits, and dZNE is calibrated on those.

To further illustrate the point that different combinations of noise factors and extrapolation functions perform best for different circuits, we systematically find the best combination for 6-qubit brickwork circuits for spin dynamics with varying two-qubit depth (0 to 40), depolarizing gate error (0.1\% to 4\%), and 8,000 shots per noise factor. The circuits are described in further detail in Appendix \ref{app:experiment}. In Fig.~\ref{fig:calibration} we plot the results as a ``calibration phase diagram". In (a), the noise-factors are 1, 3, and 5. We find that at smaller two-qubit gate depths and errors, where the noise is weak, linear extrapolation (L) with a wide range of noise factors performs the best. As the circuit area and noise increase, more and more curvature is required in the extrapolation function to extrapolate to the correct zero-noise value, thus showing preference for quadratic (Q) and exponential (E) extrapolators. We compare examples of extrapolations for select regions of these calibration diagrams in Appendix~\ref{app:extrapolation}. 

An important region of this calibration phase diagram is toward the upper-right corner, marked as \texttt{NF} (no fit) in Fig.~\ref{fig:calibration}, for large circuit areas and errors. In this limit, the depolarizing noise is completely saturated, so the expectation values of the consequent completely mixed states for traceless observables (\textit{e.g.} all Pauli operators except the identity) will converge to 0. In this regime, it is straightforward to see that linear extrapolation will always result in a low-uncertainty, zero-noise expectation value of 0 \textit{regardless of the ideal value}, which is clearly a negative outcome. Meanwhile, exponential extrapolation will be unstable, which we define as extrapolation error in the zero-noise expectation value that lies outside of the range of allowed values for the Pauli observables studied here, due to shot noise. While additional shots may diminish this instability, the overhead required scales exponentially with the noise of the target circuit, regardless of the QEM method~\citep{GoogleQEMReview2022}. 
For this experiment, we have marked a region as \texttt{NF} (no fit) if the expectation value provided by dZNE differs from the ideal expectation value by 0.4. We discuss the breakdown of extrapolation in further detail in Appendix \ref{app:extrapolation}.

\begin{figure}[ht]
\centering
\includegraphics[width=1.0\linewidth]{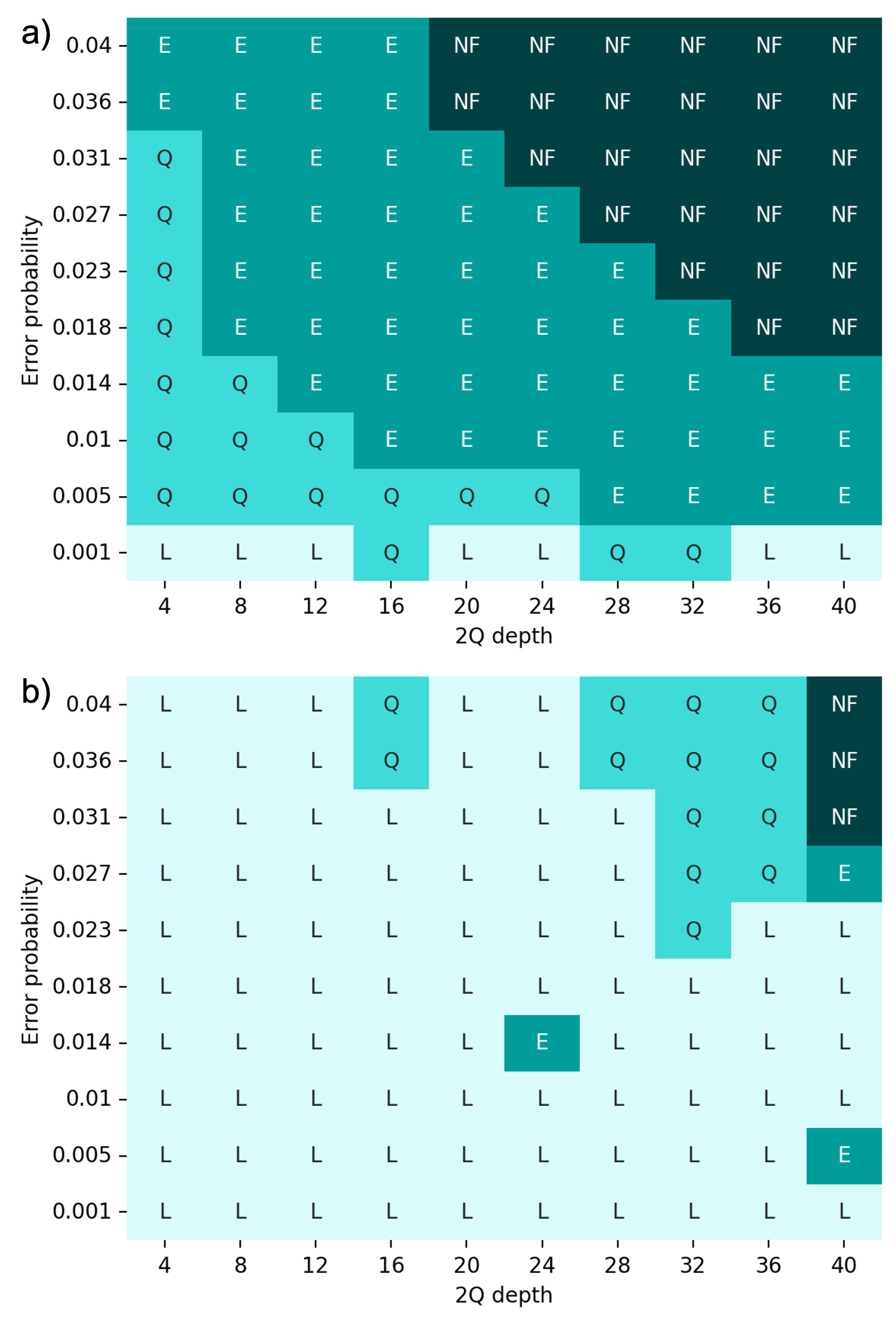}
\caption{Calibration phase diagrams for dZNE showing the best extrapolator for 6-qubit brickwork circuits with increasing two-qubit gate depth and depolarizing gate error. The regions are linear (``L"), quadratic (``Q"), and mono-exponential (``E") extrapolators, as well as ``no fit" (``NF"). The best extrapolator has the lowest root-mean-square error (RMSE) of the ideal and mitigated expectation value over multiple randomly sampled circuits with 8,000 shots per noise factor. Low-degree extrapolators are preferred unless the higher degree one can lower the RMSE by a threshold of 0.001 in this experiment. A region is said to be \texttt{NF} if the expectation value provided by dZNE differs from the ideal expectation value by 0.4. In (a), the noise factors are 1, 3, and 5, and in (b), the noise factors are 1, 1.1, and 1.2. For smaller noise factors, low-degree extrapolators are preferred for larger circuit areas and errors, and the ``NF" region is smaller.
} 
\label{fig:calibration}
\end{figure}

In Fig.~\ref{fig:calibration}(b), we compute the calibration phase diagram for a smaller range of noise factors of 1, 1.1, and 1.2. Within this smaller range of noise factors, although there exist a few anomalous entries due to random sampling, we find that the decay of the expectation value can be more appropriately described by a linear extrapolator, which is generally more stable and faster to fit, for both larger circuit areas and error rates compared to the case of noise factors of 1, 3, and 5 in Fig.~\ref{fig:calibration}(a). In addition, the ``no fit" (NF) region shrinks, suggesting that amplifying the noise to smaller factors enables the application of dZNE for deeper circuits. However, in the case of using noise factors that are not odd, positive integers, one should consider sampling multiple partially folded circuits; see Fig.~\ref{fig:partialfolding}. Finally, we note that the optimal number of noise factors, where the minimum depends on the extrapolation function, is problem-dependent and, therefore, can also be calibrated in a similar manner.

\section{Composite error mitigation strategies}

A crucial assumption of dZNE is that folding the gates amplifies \textit{all} sources of noise, enabling the decay of expectation values with increasing noise to be fitted to simple functions when the type of noise is assumed to be incoherent. In practice, both assumptions are often challenged, so we discuss ways to compose other error mitigation methods with dZNE to form \textit{composite} strategies.

\begin{figure}[ht]
\centering
\includegraphics[width=1.0\linewidth]{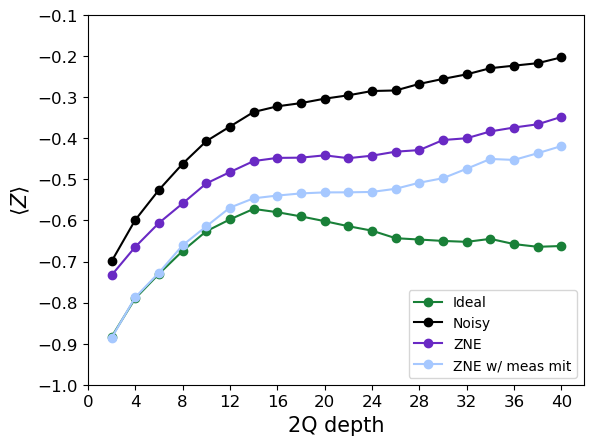}
\caption{Readout error mitigation can be straightforward concatenated with dZNE. We plot the expectation value $\langle Z \rangle = \langle Z_1 Z_2 Z_3 Z_4 Z_5 Z_6 \rangle$ with and without dZNE for a noise model with both depolarizing gate errors and readout errors. With readout mitigation, dZNE results are close to the ideal result and deviate as circuit depth increases and noisy gates dominate the total error.
} 
\label{fig:readoutmitigation}
\end{figure}

\textit{Readout error.} In conventional approaches to unitary folding, state preparation and measurement (SPAM) error is not amplified. Instead, other error mitigation techniques known to mitigate these types of errors can be injected into the dZNE workflow. In particular, we consider readout error mitigation, where it is well known that readout errors can be corrected on noisy bitstrings by measuring the readout calibration matrix \citep{Geller2020, Brayvi2021ReadoutMitigation, Nation2021mthree}, which can be simplified with twirling~\citep{vandenBerg2022TReX}, or training a model to do so \citep{Kim2022}. With the readout-corrected bitstrings, expectation values can then be computed in advance of zero-noise extrapolation. In Fig.~\ref{fig:readoutmitigation}, we demonstrate the importance of augmenting dZNE with readout error mitigation by comparing the performance of dZNE on a noisy simulator with depolarizing error and either with mitigated or unmitigated readout errors. For this experiment, we considered a depolarization noise with probability 0.01 on each \texttt{CNOT} gate, and a measurement error with $p(0|1) = 0.02$ and $p(1|0) = 0.01$; here $p(i|j)$ denotes the probability of measuring $i$ when the outcome was actually $j$. At low depths, dZNE without readout mitigation performs comparatively worse for circuits run on a noisy simulator with both depolarizing noise and readout errors, and the gap between the readout mitigated and unmitigated cases shrinks with increasing depth as gate errors become the major source of error.

\begin{figure}[ht]
\centering
\includegraphics[width=1.0\linewidth]{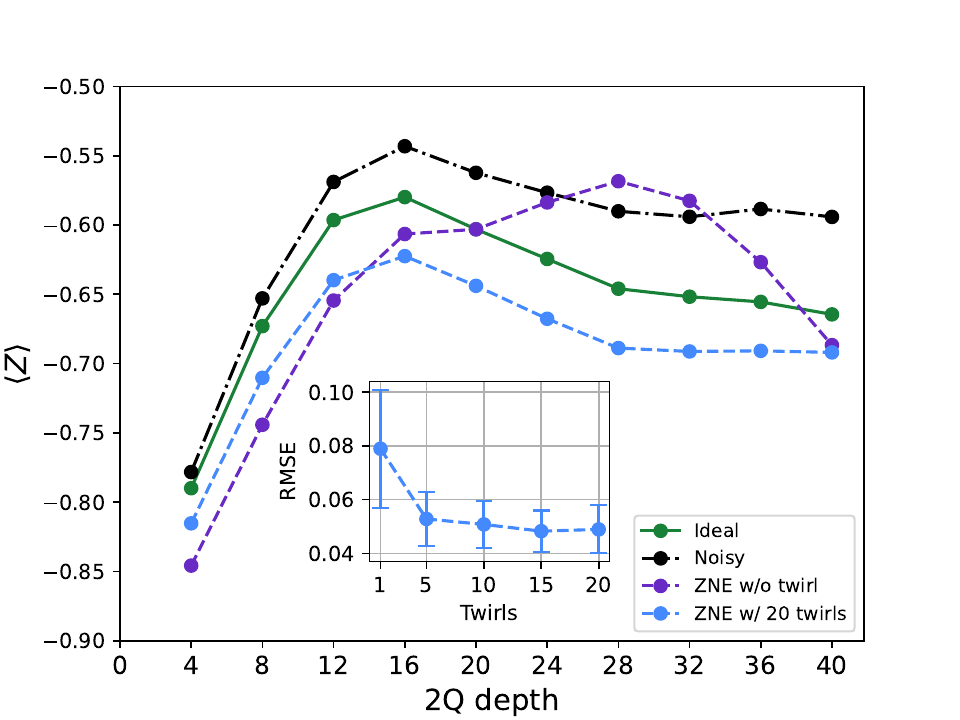}
\caption{The noise model of the device can be shaped to more closely adhere to assumptions underlying zero-noise extrapolation as a technique for quantum error mitigation. In particular, coherent gate errors can be mitigated by randomly compiling, or `twirling,' them. Here, twirling the folded 6-qubit circuits for spin dynamics results in more accurate observables. The inset shows that, for this degree of coherent gate error, the RMSE converges after sampling 10 twirled circuits.
} 
\label{fig:twirledzne}
\end{figure}

\textit{Coherent error.} Another type of error that is less straightforward to account for within dZNE is coherent error. The theory of zero-noise extrapolation with exponential functions typically assumes only the presence of \textit{incoherent} errors, so in the presence of coherent error, the analytically correct extrapolation function is generally unknown, although convergence of traceless operators to non-zero expectation values at large noise factors is a tell-tale sign of coherent error. This matter is further complicated by the ambiguity of noise amplification for gates with coherent errors, as the outcome depends on device physics-level understanding of the digital gates. Consider the case of IBM's superconducting devices, where we noted previously that the noise of the hardware-native and self-inverse \texttt{CNOT} gate can be amplified by applying more \texttt{CNOT} gates. In this case, coherent errors, say additional single-qubit rotations, could cancel out or stack depending on whether sequential \texttt{CNOT} pulses are echoed~\citep{Sheldon2016}.

Therefore, as a best practice, we suggest to randomly compile, or ``twirl," gates to remove their coherent noise. Seminal references and applications of twirling in general include Refs.~\citep{PhysRevA.94.052325, Cai2019, vandenBerg2023}, and Refs.~\citep{Kurita2022TwirlZNE, Chen2022TwirlZNE} apply twirling specifically to dZNE. Briefly, randomly compiling means that random single-qubit Pauli gates, assumed to be noiseless, are applied before and after each noisy gate in a target circuit, and averaging over multiple randomly compiled target circuits results in the asymptotic removal of coherent error. 

In Fig.~\ref{fig:twirledzne}, we compare the results of dZNE with and without twirling on a device with both depolarizing and coherent gate errors for the same circuit described in Fig.~\ref{fig:readoutmitigation}. 
Without twirling, the mitigated values are hardly improved with dZNE, but with twirling, dZNE performs much better. Theoretically, the number of possible twirled circuits scales exponentially with the number of two-qubit gates, but in practice, we notice that a modest number of twirls are sufficient for the circuit and noise model considered here to converge. For instance, in the inset of Fig.~\ref{fig:twirledzne}, we find that only 10 twirls are sufficient. Twirling also does not require a drastic overhead---although twirling requires more circuits to be run, each circuit can be run with proportionally fewer shots, ultimately conserving the shot overhead.

Finally, we note that beyond readout mitigation and twirling, it is possible to compose other error mitigation techniques, such as post-selection based on symmetries and conservation laws~\citep{Kirmani2023Braiding, Sung2023}, the insertion of dynamical decoupling pulses~\citep{Viola1999DD, Jurcevic2021, Ezzell2022DDSurvey, Pelofske2023ZNEandDD}, or virtual distillation~\citep{Bultrini2023}. 

\textit{Benchmarking on real quantum hardware.} We demonstrate that a composite error mitigation strategy, in conjunction with several ``best practices" described previously, can drastically improve results on real hardware. The composite strategy consists of readout mitigation and twirling, as well as zero-noise extrapolation with different extrapolation functions. The specific system we study is a 104-qubit 1D chain embedded in the 127-qubit cloud-hosted digital quantum computer \texttt{ibm\_washington}. We set the initial state to $|01...01\rangle$, and we apply a repeated quantum circuit similar to the many-body dynamics circuit used in Fig.~\ref{fig:calibration}-\ref{fig:twirledzne}, except that the coupling angle $\theta$ is set to 0 such that the expectation values of single-qubit $Z_i$ operators are conserved and can be straightforwardly classically computed to serve as an ideal benchmark.

\begin{figure}[ht]
\centering
\includegraphics[width=1.0\linewidth]{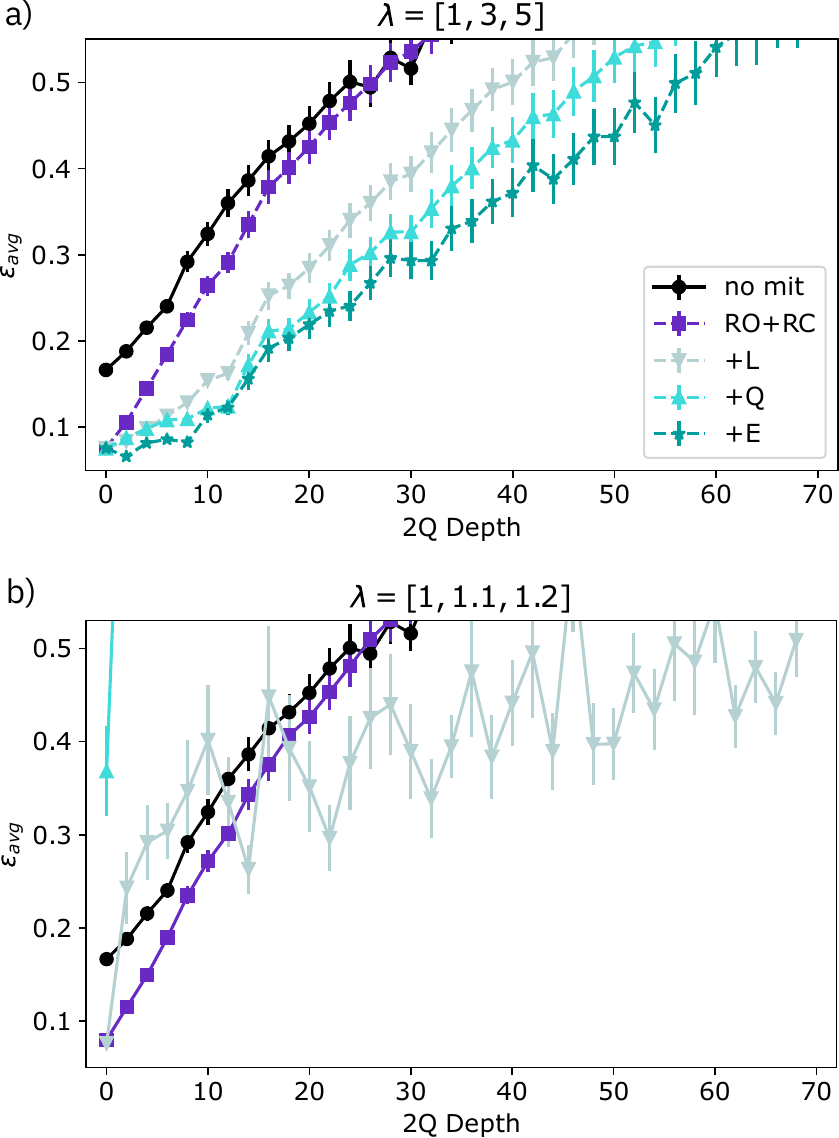}
\caption{We benchmark the proposed composite error mitigation strategy on a cloud-accessible quantum computer, \texttt{ibm\_washington}, for a 104-qubit 1D chain up to a depth of 70 two-qubit (2Q) gates for noise factors of (a) [1, 3, 5] and (b) [1, 1.1, 1.2]. For the latter, we sample only one partially folded circuit for each noise factor. We compare the cases of no mitigation (``no mit"), with readout mitigation (``RO") and randomized compiling or twirling (``RC"), and applying ZNE on top of ``RO+RC" with linear (``+L"), quadratic (``+Q"), or exponential (``+E") extrapolation. The expectation value at each noise factor is measured with 16,384 shots.
} 
\label{fig:benchmark}
\end{figure}

In Fig.~\ref{fig:benchmark}, we show the average root-mean-square error $\epsilon_\mathrm{avg}$ over all single-qubit $\langle Z_i \rangle$ up to a depth of 70 two-qubit gates, or 35 Floquet cycles. As we observed in the noisy simulations in Fig.~\ref{fig:calibration}, for noise factors of [1, 3, 5], we find that the extrapolation functions with larger curvature (\textit{i.e.} exponential or `E') produces better results at larger depths. For a smaller range of noise factors of [1, 1.1, 1.2] with a single partial fold, linear (`L') extrapolation is the only viable extrapolation function, as quadratic (`Q') and exponential extrapolations are too unstable. At the largest depths, we see that ZNE with the smaller range of noise factors [1, 1.1, 1.2] gives overall less error than ZNE with the larger range of noise factors [1, 3, 5], as the noise is less saturated in the former case. However, the depth-dependent error in the former case is less continuous, likely due to an insufficient number of partial folds. In both cases, applying readout mitigation (`RO') and twirling (`RC') improves the unmitigated results, and then applying ZNE further improves them.

\section{Conclusions and discussion}

Here, we discuss best practices for digital zero-noise extrapolation (dZNE) for QEM. During the noise amplification step, we choose local folding to enable more straightforward partial folding. We also emphasize the need to transpile circuits to specific devices before amplification to amplify the noise more predictably. During circuit execution, we highlight the importance of interleaving target and amplified circuits to minimize the impact of time-varying noise and discuss shot overhead considerations and the scaling of shot noise in the case of zero-noise extrapolation. We then provide insights into the extrapolation process and demonstrate a method of calibrating the best settings for extrapolation, such as noise factors and the form of the extrapolation function.

To address the limitations of dZNE in handling errors beyond incoherent ones, such as SPAM errors and coherent errors, we propose composite error mitigation strategies that combine dZNE with other error mitigation techniques. For SPAM errors, we recommend augmenting dZNE with readout error mitigation techniques. We highlight the importance of this approach by comparing the performance of dZNE on a noisy simulator with depolarizing error and readout errors. We then discuss the challenges in accounting for coherent errors within the dZNE framework, particularly due to the ambiguity of noise amplification for gates with coherent errors. As a best practice, we suggest randomly compiling, or ``twirling," gates to remove their coherent errors.

In the future, we expect studies that extend these best practices to more specific contexts, such as particular circuit classes, device noise profiles, or gate sets, to be fruitful. These studies could provide a deeper understanding of the method's limitations and suggest improvements to its robustness. We also anticipate that further development of dZNE will re-define best practices. Specific approaches that warrant further exploration include adaptive dZNE for dynamically allocating shots between circuits with a range of noise factors, and machine learning and data-driven approaches for optimizing dZNE parameters, such as noise factors, extrapolation functions, and folding techniques. 

\section*{Acknowledgements}
We acknowledge Mirko Amico, Paul Nation, Zlatko Minev, and Sanket Panda for valuable discussions. 
This work was supported by the Swiss State Secretariat for Education, Research, and Innovation (SERI) under contract number MB22.00051.

\appendices

\section{Extrapolation} \label{app:extrapolation}

\begin{figure*}[htb]
\centering
\includegraphics[width=\textwidth]{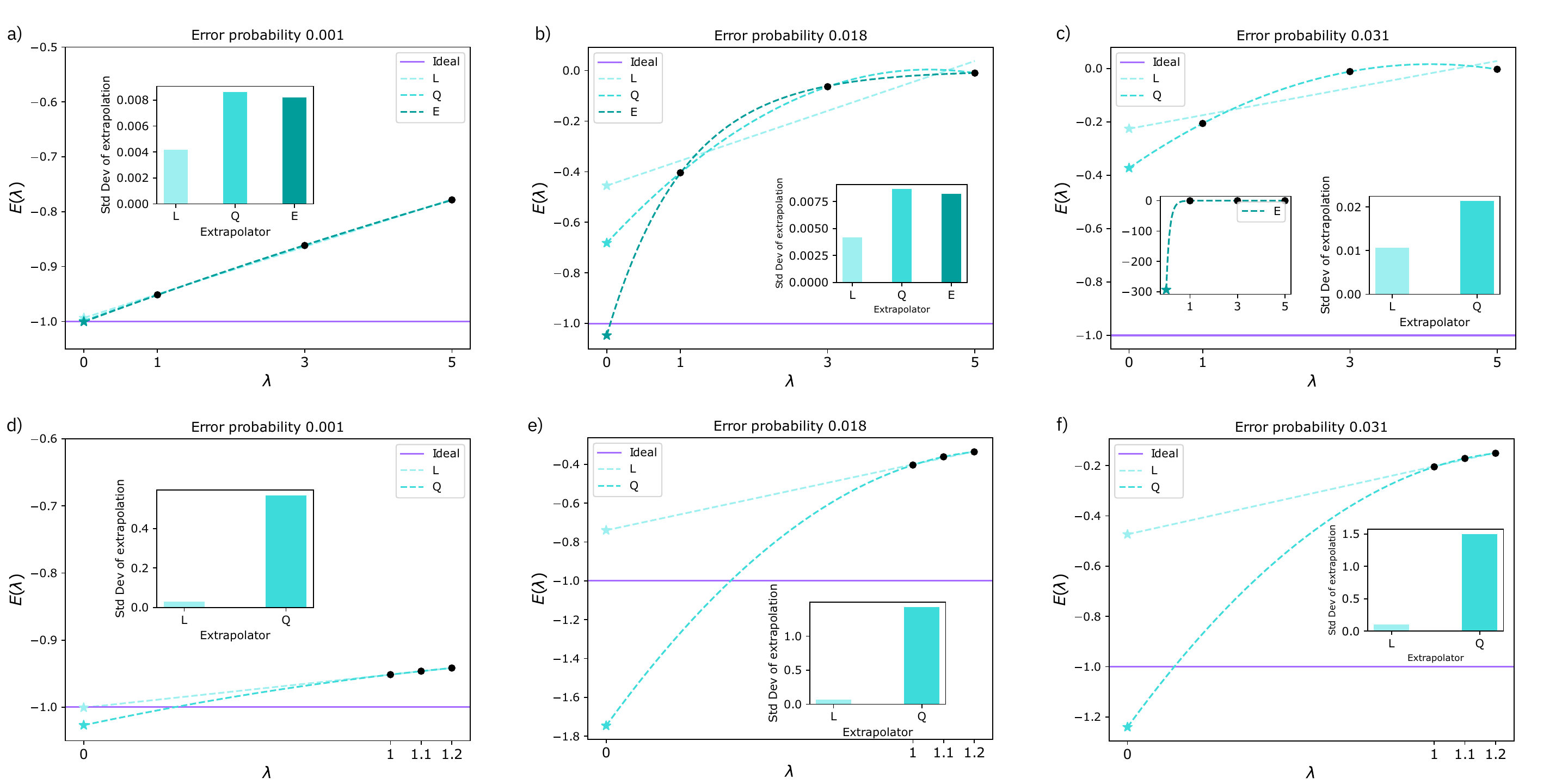}
\caption{Extrapolation for different regions of the calibration phase diagrams in Fig.~\ref{fig:calibration}. The top (bottom) row is for noise factors of [1, 3, 5] ([1, 1.1, 1.2]), and the depolarizing gate error probability increases from left to right. As the error increases, extrapolation functions with more curvature (\textit{e.g.} exponential) are required to minimize the bias compared to an extrapolation function with less curvature (\textit{e.g.} linear). However, such functions have more variance in the extrapolated value. Additionally, smaller noise factors enable ZNE to be used for larger circuit areas.}
\label{fig:expval}
\end{figure*}

Here, we develop further understanding of the trends observed in the calibration phase diagrams of Fig.~\ref{fig:calibration} by analyzing extrapolation plots in Fig.~\ref{fig:expval}. These trends include that low-order extrapolation functions are preferred at low 2Q depth, low error probability, and smaller ranges of noise factors, as well as that smaller ranges of noise factors enable lower errors at large circuit areas. In particular, we show the zero-noise expectation values from different extrapolators for noise factors [1, 3, 5] (top row) and [1, 1.1, 1.2] (bottom row). We choose three probabilities of error such that they correspond to regions `L', `E' and `NF' of Fig.~\ref{fig:calibration}(a).

First, we discuss noise factors [1, 3, 5] (top row). For low probability of error (0.001) on the left, all three extrapolators have comparable performance in terms of the extrapolated value, with the linear extrapolator having the smallest standard deviation. This changes as the error probability is increased. For error probability 0.018, exponential extrapolator is the best, although it has a higher standard deviation than both linear and quadratic. We also note that the standard deviation for all the extrapolators increases with increasing noise. Finally, for error probability 0.031, none of the extrapolators are close to the ideal expectation value; the exponential extrapolator drifts off to a totally wrong value. This region corresponds to 'NF' in Fig.~\ref{fig:calibration} (a).

For noise factors [1, 1.1, 1.2] (bottom row), i.e., partial folding, we see that the linear extrapolator remains the best for both error probabilities 0.001 and 0.018. Furthermore, for error probability 0.031, where for noise factors [1, 3, 5] no fit was obtained, the quadratic extrapolator is able to reach a value close to the ideal expectation value. This re-affirms that partial folding to a smaller range of noise factors enables the application of ZNE up to a higher error rate. However, a drawback is that the standard deviation of all the extrapolators is much higher for partial folding, thus requiring multiple samples of partially folded circuits as discussed in Section~\ref{noiseamplification}.

As we show in Section~\ref{sec:extrapolation}, when the circuit area and error become too large, the exponential extrapolator often fails to converge on a reasonable estimate for the zero-noise value. This failure is due to numerical instabilities that make such extrapolators difficult to fit~\citep{Bi2022StabilizationOP}. Here, we describe the onset of these numerical instabilities and what can be done to detect and correct for them in practice.

For a general quantum observable $\hat{O}$, the exponential extrapolator is often the most natural for ZNE \citep{Cai2020MultiexponentialEE}. For simplicity, we assume a mono-exponential form for the decay of the expectation value: $\langle \hat{O} \rangle = S + A e^{-i\lambda/L}$. Three degrees of freedom are required to fit this curve as a function of the noise $\lambda$: the shift coefficient $S$, the amplitude $A$, and the decay length $L$. 

The exponential model is intrinsically unstable when fitting to expectation values that are $\lambda$-independent within estimation error including, for instance, shot noise. Such results can occur either when the noise has saturated ($\lambda_\mathrm{min} \gg L$) or is too low ($\lambda_\mathrm{min} \ll L$), and both situations can lead to possibly misleading results without properly accounting for this instability. The latter situation, however, is comparatively benign---one can simply decide not to use ZNE at all or switch to a linear extrapolator. The former situation can be especially misleading because instabilities can manifest in multiple ways, depending on the noise model and observable. For instance, in the case where there is coherent error or the observable is not traceless, such that $S$ may not be 0, then there is ambiguity in fitting $S$ and $A$. Even when $S$ is known, there is ambiguity in $L$ and $A$, since any small $L \ll \lambda_\mathrm{min}$ could result in expectation values that are virtually $\lambda$-independent. In both cases, because the correct value of $A$ is ambiguous, the zero-noise expectation value is unstable. 

While detecting and correcting instabilities is likely context-dependent, we suggest some general strategies. After extrapolation, we recommend comparing extrapolation functions, \textit{e.g.} linear, quadratic, and exponential,  and choosing the result based on metrics such as the standard error of the extrapolated value or goodness of fit. Before extrapolation, one should determine whether expectation values vary sufficiently with the degree of noise amplication, or whether the noise is too high or low for stable extrapolation. We anticipate the development of formal protocols for choosing the extrapolation function and noise factors to be a valuable contribution.

\section{Experimental details} \label{app:experiment}

\subsection{Circuits for spin dynamics}

\begin{figure}
    \centering
    \begin{adjustbox}{scale=0.7}
    \begin{quantikz}
    {q_0}& \qw & \ctrl{1}\gategroup[wires=6, steps
    =5, style={dashed,rounded corners,inner sep=6pt}]{} & \gate{U_3 (\theta_1, \theta_2, \theta_3)} & \gate{P(\phi_1)} & \qw & \qw & \qw\\
    {q_1}& \gate{X} & \control{} & \gate{U_3 (\theta_1, \theta_2, \theta_3)} & \ctrl{1} & \gate{U_3 (\theta_1, \theta_2, \theta_3)} & \gate{P(\phi_2)} & \qw\\
    {q_2}& \qw & \ctrl{1} & \gate{U_3 (\theta_1, \theta_2, \theta_3)} & \control{} & \gate{U_3 (\theta_1, \theta_2, \theta_3)} & \gate{P(\phi_3)} & \qw\\
    {q_3}& \gate{X} & \control{} & \gate{U_3 (\theta_1, \theta_2, \theta_3)} & \ctrl{1} & \gate{U_3 (\theta_1, \theta_2, \theta_3)} & \gate{P(\phi_4)} & \qw\\
    {q_4}& \qw & \ctrl{1} & \gate{U_3 (\theta_1, \theta_2, \theta_3)} & \control{} & \gate{U_3 (\theta_1, \theta_2, \theta_3)} & \gate{P(\phi_5)} & \qw\\
    {q_5}& \gate{X} & \control{} & \gate{U_3 (\theta_1, \theta_2, \theta_3)} & \gate{P(\phi_6)} & \qw & \qw & \qw
    \end{quantikz}
    \end{adjustbox}
    \caption{A 6-qubit circuit for spin dynamics. The boxed region denotes one time-evolution step with a two-qubit depth of two that can be repeated to increase the depth of the circuit. This figure was drawn using \textit{Quantikz} \citep{kay2018tutorial}.}
    \label{fig:circuit}
\end{figure}
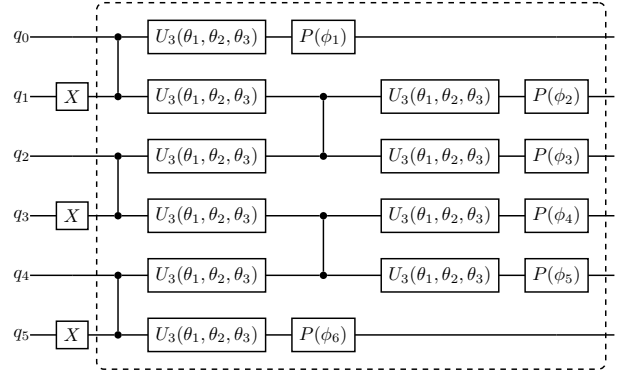

The results in Fig.~\ref{fig:calibration},~\ref{fig:readoutmitigation}, and~\ref{fig:twirledzne} were performed on brickwork circuits for spin dynamics as shown in Fig.~\ref{fig:circuit}. These circuits are parametrized by disorder $\phi_i$ for each qubit $i$ and interaction strength $\theta_1$. We set $\theta_2=\theta_3=0$. The disorders are randomly sampled from a uniform distribution between $-\pi$ and $\pi$, and $\theta_1$ is set to $0.05 \pi$, such that local charges do not decay too quickly. Each time-evolution step in the dotted box in Fig.~\ref{fig:circuit} has a two-qubit depth of 2. In our calculations, we study circuits with up to 35 steps, or two-qubit depth of 70. For the 104-qubit example run on real quantum hardware, the largest circuit contains $3,605$ two-qubit gates.

\subsection{Software implementation}

For this study we have used a number of open-source tools. Zero-noise extrapolation was performed using the open-source \texttt{prototype-zne}~\citep{Rivero2022ZNE}, which wraps the \texttt{Estimator} primitive as defined in \texttt{Qiskit}'s standard \texttt{BaseEstimator} interface \citep{Qiskit}. Much of the functionality discussed is also available to users via IBM's Qiskit Runtime. 

\end{document}